# Micromegas in a Bulk

*I. Giomataris[1], R. De Oliveira[2], S. Andriamonje[1], S. Aune[1], G. Charpak[2], P. Colas[1], A. Giganon[1], Ph. Rebourgeard[1], P. Salin[3]*

[1]CEA-DSM/Saclay-France
[2]CERN, Geneva, Switzerland
[3]PCC-CdF-IN2P3-France

**Abstract**

In this paper we present a novel way to manufacture the bulk Micromegas detector. A simple process based on the PCB (Printed Circuit Board) technology is employed to produce the entire sensitive detector. Such fabrication process could be extended to very large area detectors made by the industry. The low cost fabrication together with the robustness of the electrode materials will make it extremely attractive for several applications ranging from particle physics and astrophysics to medicine.

## Introduction

The driving factor in the development of new gaseous detectors comes from the advent of new accelerators, especially LHC (Large Hadron Collider). They required a capacity of resisting to higher rates than the wire chambers and providing a superior position and time resolution in order to meet the additional requirements specific to each accelerator conditions. An important objective will be a drastic reduction of the material budget and dead regions (for instance due to mechanical frames). The decision to take a safer path, be it at a much-increased price and with several drawbacks like a larger mass was made for the LHC inner tracking detectors because at that time no gaseous detector could prove to be a totally safe solution.

After four years of intense research it is now clear that various gaseous detectors can now match the requirements for a central tracking detector at a much lower cost than solid state detectors and with better characteristics in terms of time resolution, amount of material and cost.

Among the new innovating detectors the Micromegas [1] approach has now reached maturity and is suceesfully used by many experiments like COMPASS, NA48, CAST, n-TOF and it under study for the future Linear Collider [2-6]. The detector is also under development for low energy neutrino experiments (HELLAZ, NOSTOS) [7-8], including coherent neutrino scattering, neutrino magnetic moment measurements and dark matter searches.

The design of Micromegas is offering substantial advantage in counting rate, energy, spatial and time resolution, granularity on large surface and simplicity [9-11]. The amplification gap of Micromegas is obtained by suspending a mesh over the anode strips or pads. The precise gap, which is narrow, usually 50-150 μm is obtained by using adequate insulating spacers (pillars) printed on top of the anode plane by conventional lithography of a photoresistive film. The mesh is stretched and glued on a frame and then rested on top of the pillars. The challenge to the technician is on the handling of the mesh to obtain a rather good flatness and parallelism between the anode and cathode (mesh); then applying a voltage on both sides of the gap, the intense electric field pulls down the mesh and the flatness is then defined by the height of the pillars which have an accuracy of better than 10 μm.

One important factor in the delay of the adoption of the new gaseous detectors is the fact that the technology to build them is not straightforward and the spread in their use was very much influenced by the capacity of large laboratories like CERN or Saclay to help by providing good prototypes.

We present here a new method of producing large area Micromegas elements in one single process that will make the mounting of the detector a simple game. Moreover the technology used is robust and the probability of damaging the detector during mounting and testing is very small.

## Fabrication process

In order to obtain large area detectors and more reliable structure we decided to use a woven wire mesh, instead of the usual electroformed micromesh, as has been suggested and used for a low background TPC (12). There are several advantages to use these quite conventional meshes:
1.  they exist in rolls of 4 m x 40 m and are quite inexpensive,
2.  they are commonly produced by several companies over the world,
3.  there are many metals available: Fe, Cu, Ti, Ni, Au,
4.  they are more robust for stretching and handling.

For our first prototype we have used a stainless-steel cloth mesh made by stainless steel wires of 30 μm in diameter interwoven at a pitch of 80 μm.

We adopted a simple technical solution to build the whole detector in one process: the anode plane carry the copper strips, a photo resistive film having the right thickness and the cloth mesh are laminated together at high temperature, forming a single object. By photolithographic method then the photo resistive material is etched producing the pillars. The pillars have cylindrical shape of 400 μm in diameter and were printed with a distance of 2 mm as shown in Fiogure 1. With this method we have constructed two prototypes, one having a amplification gap of 75 μm and the second of 150 μm, both with a square active area of 9x9 cm$^2$.

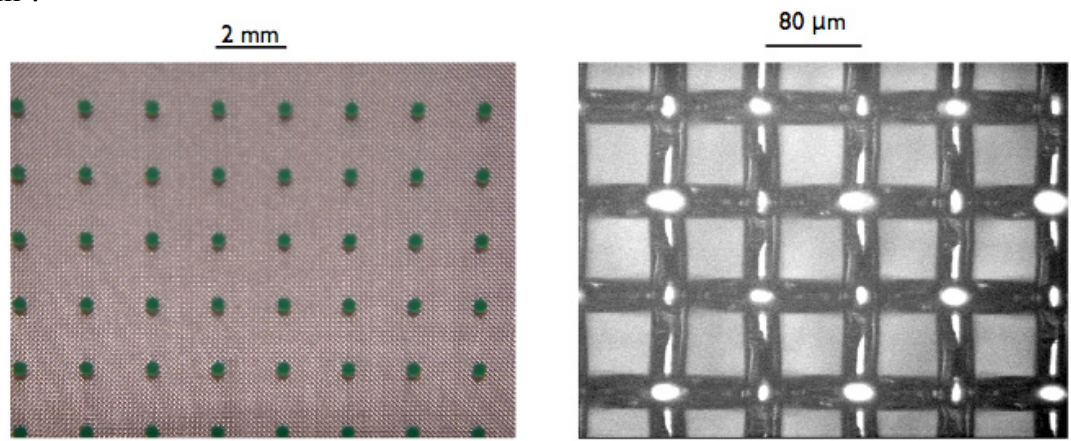

Figure 1

Photographs of the bulk detector elements. The picture at left shows a small area of the detector; the 400 μm in diameter pillars every 2 mm are visible. On the right side is a microscopic view showing details of the woven wire mesh.

The test chamber was introduced in a gas vessel and the experimental setup used for the test is shown in Figure 2. It consists of the bulk detector surrounded by a 10 mm drift region defined by a simple drift electrode made also by a cloth grid. The gas filling was a mixture of Argon (95%) and isobutane (5%).

The primary charges were produced in the drift volume by an X-ray $^{55}$Fe source and then transported through the electric field to the bulk element and amplified in the small amplification gap of the Micromegas detector. Fast signals are induced in both anode and cathode plane. We have used a low noise charge preamplifier to read-out the cathode-mesh signal. As expected the rise time of the signal, due to the ions produced during the avalanche process, was quite fast, 60 ns for the 75-μm gap and 150 ns for the 150-μm gap.

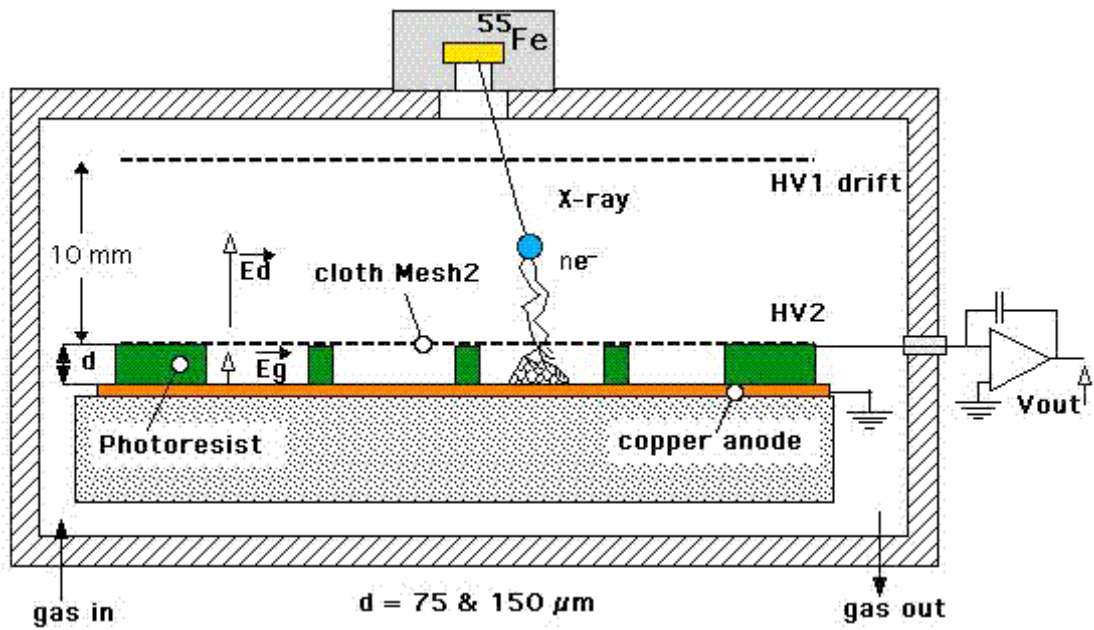

Figure 2 - Schematic drawing of the test chamber

The gain obtained was quite satisfactory for both amplification gaps reaching a maximal value, before break down, of about $2 \times 10^4$, Figure 3 shows the measured gain as function of the applied voltage for the two bulk detectors. Applied voltages are quite reasonable ranging from 300-400 Volts and 400-500 Volts for respectively the 75 and 150-μm amplification gap.

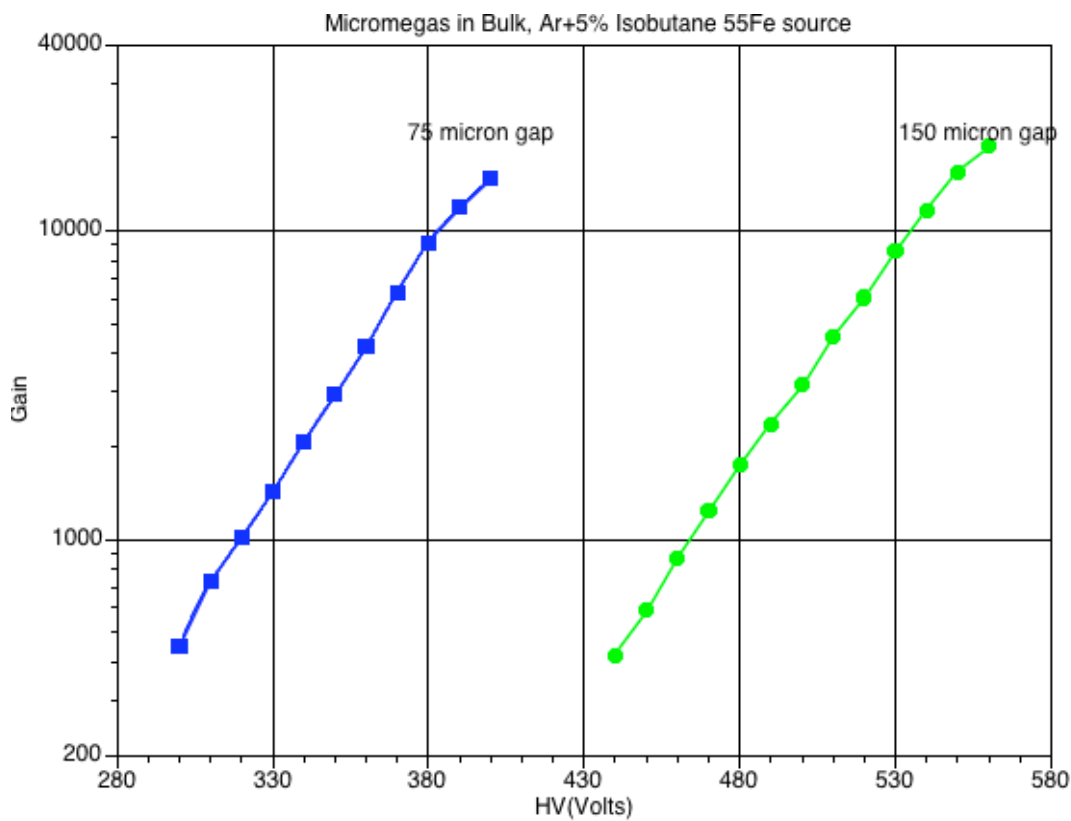

Figure 3: Gain curves obtained by the two detectors

The energy resolution was measured by the pulse height distribution of the collected signal in the mesh, proportional to the deposited energy. As shown in figure 4 the energy resolution is quite reasonable, about 20% (FWHM) for the 5.9 keV deposited energy. To reach the same energy resolution, as in the case of a conventional Micromegas equipped with a thin mesh, we need a thinner cloth mesh thickness (smaller wires) and laminated.

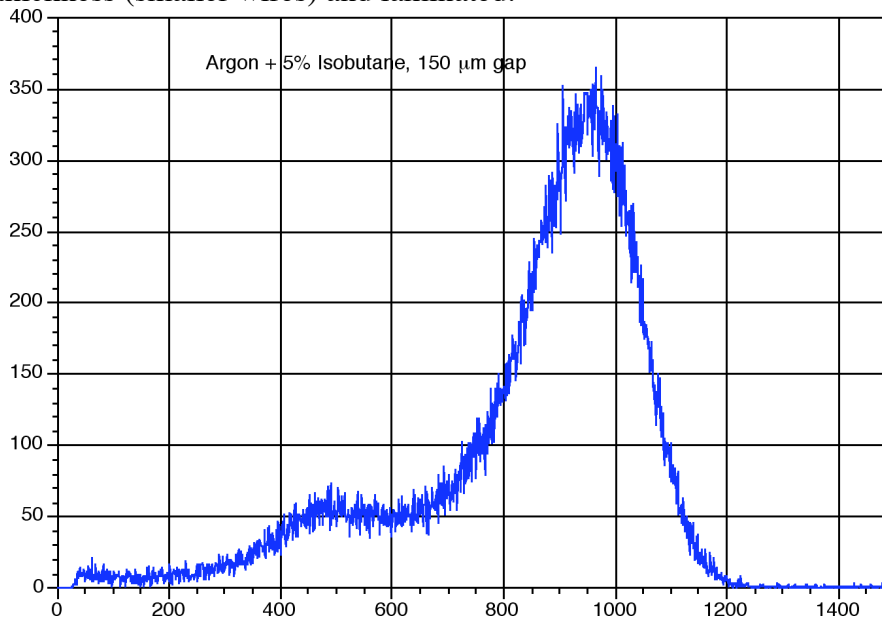

Figure 4: Energy resolution of the 150-µm gap detector, measured with $^{55}$Fe

5.9 keV X-rays.

**Future improvements**

To benefit from this progress we propose to perfect this technology for use in low background astroparticle physics experiments that require large volume gas detectors. Thus we aim with this project to focus on two major objectives:
- A significant improvement will be the use of a thinner and flatter mesh. In order to obtain a smooth surface and a lower thickness, we will employ an industrial standard process that consists of passing the cloth mesh through a pair of heavy rolls to decrease the thickness and flatten the intersections. We expect an improvement of the energy resolution and an easier passage of the drifting electrons.
- Build large area (several m$^2$) by low cost industrial process.
- Develop by the same technology double-stage amplification detectors that can be useful to cope with very-high intensity hadron beams or operation in high pressure Xenon gas mixtures, as it is required in medical applications.
- Apply the same procedure on a flexible anode board (for instance a thin Kapton). It will allow a further reduction of the material budget of the detector and open the way for a curved Micromegas.
- In the present study we took too much margin factor for the diameter and pitch of the spacers. We will push the technology as much as possible to obtain larger pitch and smaller diameter of the pillars in order to reduce the detector dead space to a negligible level.

The improvement of the Micromegas technology will bring finally the problem of selection of one type or another to a realistic estimation of the cost, of the difficulty of construction and the physical characteristics.
It will sometimes depend on the type of experiment one wish to undertake, where the differents approaches to micropattern gaseous detectors have specific advantages.
But it seems to us reasonable to think that the simplicity of the construction, the low material budget and cost together with the high spatial and time accuracy of the Micromegas may cover areas where the solid-state detector has been used.